\documentstyle[12pt]{article}

\def\p{{\rm {\bf p}}}

\begin{document}

\noindent Yukawa Institute Kyoto

\hfill YITP/K-1075

\hfill June 1994

\vspace{1.5cm}

\begin{center}
{\bf THE SPECTRAL DENSITY METHOD APPLIED TO THE HOT QUANTUM
FIELD THEORIES}

\vspace{1.5cm}
{\bf O.K.Kalashnikov}
\footnote{Permanent address: Department of
Theoretical Physics, P.N.Lebedev Physical Institute, Russian
Academy of Sciences, 117924 Moscow, Russia. E-mail address:
kalash@td.fian.free.net}

Yukawa Institute for Theoretical Physics

Kyoto University,

Kyoto 606-01, Japan
\vspace{2.5cm}

{\bf ABSTRACT}

\end{center}

The spectral density method being applied to the quantum field
theory at finite temperature is revived and its possibilities
are briefly discussed.

\newpage

\section {Introduction}

The spectral density method proposed for the first time in 1968
[1] is surveyed and we find it to be useful for studying the hot
gauge theory and other relativistic models. This method is shown
through two models but one can see that its main content keeps
the same as was done previously (see e.g. Ref.[2] for details).
We demonstrate that its simplest approximation is very convenient
to fix the nontrivial vacuum and to build a reliable
calculational scheme on this basis. To date, the separate
fragments of this method have been already used by many authors
for different applications (see e.g. [3,4] and [5]) although its
complete formalism remains unknown. The goal of this note is to
represent this formalism and to show that no peculiarities arise
when its formulae are applied to the relativistic models which
are intensively studied now but (unfortunately) they were not
considered in [2]. Below, this method is briefly introduced in
accordance with the paper [2] and its simplest approximation is
discussed.

\section {The general formalism}

The spectral density is determined as a standard statistical
average of a non-equal-time commutator or anticommutator ($\eta=
\mp 1$) of some operators A and B:
\setcounter{equation}{0}
\begin{eqnarray}
\Lambda(\omega)=\eta <[A;B(\tau)]_\eta>_\omega=
\int d\tau exp(i\omega\tau)
(\eta[A;B(\tau)]_\eta>)
\end{eqnarray}
and should be found through the exact set of integral relations
\begin{eqnarray}
\int\frac{d\omega}{2\pi}\omega^m(\eta<[A;B(\tau)]_\eta>)_\omega
=\big<\left[[H[H...[H;A]_-...];B\right]_\eta\big>
\end{eqnarray}
which are a direct concequence of the equation of motion for the
operator $B(\tau)$
\begin{eqnarray}
i\frac{\partial}{\partial\tau}B(\tau)=[B(\tau);H(\tau)]_-
\end{eqnarray}
This is the essence of the spectral density method and the
connection with the standard Green function formalism is
established through the following formula
\begin{eqnarray}
{\cal D}(\omega)=\int\frac{dz}{2\pi}
\frac{\Lambda(z)}{z-\omega}
\end{eqnarray}
where the standard analytical conditions should be taken into
account. The iteration ansatz within Eq.(2) and another useful
formulae can be found in [2] but it is not our goal to rewrite
them again. In what follows we shall try to demonstrate this
method and discuss its possibilities.

\section {U(1)-scalar field}

This is the simplest case for testing our method. The Hamiltonian
is defined to be
\begin{eqnarray}
H=\frac{1}{2V}\sum_\p\left[\pi_{\p}\pi_{-\p}+(\p^2+m^2)
\varphi_{\p}\varphi_{-\p}\right]
\end{eqnarray}
where the system is put into the three dimentional black box V
with the periodic boundary conditions. This is a useful way
to simplify the standard commutation relations and to avoid
the needless difficulties. Now all moments for the standard
spectral density
\begin{eqnarray}
\Lambda(\p,\omega)=-<[\varphi_\p(0);\varphi_\p(\tau)]_->_\omega
\end{eqnarray}
are easily calculated and found to be
\begin{eqnarray}
\int\frac{d\omega}{2\pi}\omega\Lambda(\p,\omega)=1\,,\qquad
\int\frac{d\omega}{2\pi}\omega^{2n+1}\Lambda(\p,\omega)
=(\p^2+m^2)^n
\end{eqnarray}
where $n=1,2$ and so on. Only the odd moments are not equal zero
and this is the general situation for all relativistic systems.
The exact spectral density has the form
\begin{eqnarray}
\Lambda(\p,\omega)=\frac{\pi}{\omega(\p)}[\,\delta(\omega-
\omega(\p))-\delta(\omega+\omega(\p))\,]
\end{eqnarray}
and all moments will be satisfied if $\omega(\p)=\sqrt{\p^2
+m^2}$.

\section{SU(N)-gauge theory}

The first main question is to choose the gauge and to define the
Hamiltonian in the most simple way. This situation is typical
for any calculations within a gauge theory and should be solved
at the beginning in the standard manner. To this end the temporal
axial gauge is fixed where the Hamiltonian is known to be
\begin{eqnarray}
H=\frac{1}{2}\pi_i^a\pi_i^a+\frac{1}{4}B_i^aB_i^a\,\,,\nonumber\\
B_i^a=\partial_jV_l^a-\partial_lV_j^a+gf^{abc}V_j^bV_l^c
\end{eqnarray}
and the usual gluon propagator has the form
\begin{eqnarray}
{\cal D}_{ij}(p_4,\p)=\frac{1}{(p_4^2+\p^2)+A(p)}
\left(\delta_{ij}-\frac{p_ip_j}{\p^2}\right)
+\frac{1}{p_4^2+(p_4^2/\p^2)\Pi_{44}(p)}\frac{p_ip_j}{\p^2}
\end{eqnarray}
Both scalar functions $A(p_4,\p)$ and $\Pi_{44}(p_4,\p)$ have
the completely different limits in accordance with the ratio
$p_4/|\p|$ and here only the so-called "plasmon limit" (but not
the infrared one) is essential. This limit for the small momenta
was calculated many years ago [6] and known to be
\begin{eqnarray}
A(p_4,\p)=m^2\left(1-\frac{\p^2}{5p_4}\right),\,\qquad
\frac{p_4^2}{\p^2}\Pi_{44}(p_4,\p^2)=m^2\left(1-
\frac{3\p^2}{5p_4^2}\right)
\end{eqnarray}
where $m^2=g^2T^2N/9$. The standard spectra
\begin{eqnarray}
\omega_t^2(\p)=m^2+\frac{6}{5}\p^2,\,\qquad
\omega_l^2(\p)=m^2+\frac{3}{5}\p^2
\end{eqnarray}
result from the additional iteration within (10) when $p_4^2$
replaces to $(-m^2)$ for Eq.(11). This means that the
simplest approximation within the spectral density method should
reproduce another spectra which ignore the last iteration but
(as we shall see) give the qualitatively correct result.

The correspondent spectral density is defined to be
\begin{eqnarray}
\Lambda_{i,j}(\p,\omega)=-<[V_{i,\p}(0);V_{j,\p}(\tau)]_->\omega
\end{eqnarray}
and has a standard tensor stucture
\begin{eqnarray}
\Lambda_{i,j}(\p,\omega)=\left(\delta_{ij}-
\frac{p_ip_j}{\p^2}\right)\Lambda_t(\p,\omega)
+\frac{p_ip_j}{\p^2}\Lambda_l(\p,\omega)
\end{eqnarray}
Its first nonzero moments are easily calculated
\begin{eqnarray}
\int\frac{d\omega}{2\pi}\omega\Lambda_{i,j}(\p,\omega)=
\delta_{i,j}\,,\qquad
\int\frac{d\omega}{2\pi}\omega^3\Lambda_{i,j}(\p,\omega)
=\left[(\p^2+m^2)\delta_{i,j}-p_ip_j\right]
\end{eqnarray}
and allow to find all parameters if the spectral density has the
stadard form
\begin{eqnarray}
\Lambda_n(\p,\omega)=\frac{\pi}{\omega_n(\p)}[\,\delta(\omega-
\omega_n(\p))-\delta(\omega+\omega_n(\p))\,]
\end{eqnarray}
where $n=t,l.$ The result is
\begin{eqnarray}
\omega_t^2(\p)=\p^2+m^2,\,\qquad\omega_l^2(\p)=m^2
\end{eqnarray}
and one can see that this is qualitatively the same as Eq.(12).
But within the spectral density method we can find a
selfconsistent equation for $m^2$ which is obtained in (15) to be
\begin{eqnarray}
\delta_{i,j}m^2\delta^{a,b}=-\frac{1}{2}\Gamma_{ijmn}^{abcd}
\int\frac{d^3\p}{(2\pi)^3}<V_{m,\p}^cV_{n,-\p}^d>
\end{eqnarray}
after the third moment has been calculated. Using the known
formula from Ref.[2]
\begin{eqnarray}
<A\,B>-\theta(-\eta)<A><B>=\int\frac{d\omega}{2\pi}
\frac{(\eta<[A,B(\tau)]_\eta>_\omega)}{exp(\beta\omega)+\eta}
\end{eqnarray}
and the standard dimensional regularization one find this
equation to be
\begin{eqnarray}
m^2=\frac{4g^2N}{3}\int\frac{d^3\p}{(2\pi)^3}\frac{1}
{\sqrt{m^2+\p^2}}\frac{1}{exp{[\beta\sqrt{m^2+\p^2}]}-1}
\end{eqnarray}
which reproduces the known value $m^2=g^2T^2N/9$ in the hight
temperature limit. The more complicated approximations for
$\Lambda_{i,j}(\omega)$ are also available (see Ref.[2]) to
obtain Eq.(12) and other peculiarities of the SU(N)-gauge
theory but it is not subject of this note.

\section {Conclusion}

Here we demonstrate that the spectral density method (being very
simple) is able to give the qualitatively correct spectra (and
other quantities) for any quantum system and can be used to
build a reliable calculational scheme beyond the standard
perturbative expansion.

\newpage

\section {Acknowledgements}

I would like to thanks all the colleagues from the Yukawa
Institute for the useful discussions and the kind hospitality.
Especially I am very obliged to Professor Y. Nagaoka for
inviting me to the Yukawa Institute where this investigation
has been performed.

\section {References}

\renewcommand{\labelenumi}{\arabic{enumi}.)}
\begin{enumerate}

\item{ O.K.Kalashnikov and E.S.Fradkin, Zh. Eksper. Teor. Fiz.
{\bf 55} (1968) 607 [Sov. Phys. - J. Exper. Theor. Phys. {\bf
28} (1969) 317].}

\item{ O.K.Kalashnikov and E.S.Fradkin, Phys. Stat. Sol. (b)
{\bf 59} (1973) 9. }

\item{ R.D.Pisarski, Physica {\bf A158} (1989) 146.}

\item{ E.Braaten and R.D.Pisarski, Nucl. Phys. {\bf B337} (1990)
569; Nucl. Phys. {\bf B339} (1990) 310;  E.Braaten and T.C.Yuan,
Phys. Rev. Lett.{\bf 66} (1991) 2183.}

\item{ H.Schulz, Phys. Lett. {\bf B291} (1992) 448.}

\item{ O.K.Kalashnikov V.V.Klimov, Yad. Fiz. {\bf 31} (1980)
1357 [Sov. J. Nucl. Phys. {\bf 31} (1980) 699]. }

\end{enumerate}

\end{document}